\documentclass[prd,twocolumn,amsmath,amssymb,nofootinbib]{revtex4}
\usepackage{graphicx}
\usepackage{dcolumn}
\usepackage{bm}
\usepackage{epsfig}

\begin{document}
\title{Pentaquarks in the Chiral Symmetry Limit} 
\author{Dmitri Melikhov$^{a}$ and Berthold Stech$^{b}$}
\affiliation{
$^a$ Nuclear Physics Institute, Moscow State University, 119992, Moscow, Russia\\
$^b$ Institut f\"ur Theoretische Physik, Universit\"at Heidelberg, 
Philosophenweg 16, 69120, Heidelberg, Germany}
\begin{abstract}
We demonstrate that a five quark state of positive parity with an
internal $P$-wave structure  - usually pictured as a composite of 
an antiquark and two diquarks in a $P$-wave state - can couple to
nucleons and Goldstone particles in a chirally invariant way. 
The corresponding decay width is generally not suppressed. 
A pentaquark of positive or negative parity with an internal $S$-wave 
structure, which may be viewed as a composite of an antiquark and 
two chirally different diquarks in an $S$-state, does not couple to 
nucleons and light mesons in the limit of an unbroken chiral
symmetry. It is stable in this limit. However, such states can decay
via the  effect of the spontaneous breaking of chiral symmetry.  This
breaking is strong because of the sizeable magnitude of the quark
condensate. 
Thus, chiral symmetry cannot be the cause of a tiny decay amplitude, 
even for pentaquarks stable in a strict chiral symmetry limit.
\end{abstract}
\maketitle
\section{introduction}
The existence of pentaquarks \cite{signal} is not yet undoubtedly established. 
But if these particles  exist the exotic members of the pentaquark
multiplet must have a very small decay width of order 1 MeV or even 
lower \cite{sibirtsev}. For the internal structure of the pentaquark most likely
diquarks play an important role \cite{jw}. In fact, quark-quark 
correlations are essential in low energy processes and provided e.g.  
the explanation for the huge $\Delta I= 1/2$ enhancement in 
weak decays \cite{neubert}. For the possible origin of the small width 
of the pentaquark many qualitative suggestions have been put forward. 
In a scenario proposed by Jaffe and Wilczek \cite{jw} the pentaquark 
consists of an antiquark and two scalar diquarks a in a relative 
$P$-wave angular momentum state.
However, a fully dynamical quark model calculation using a non
relativistic Fock space representation for the pentaquark $\Theta^+$ 
in the Jaffe-Wilczek scenario showed, that the color and 
flavor factors do not reduce the width sufficiently.   
A small spatial overlap of the wave functions is necessary which 
requires an uncommon spatial structure of this pentaquark \cite{mss}. 

Recently, the chiral symmetry of QCD was considered as another
possible cause of the small width \cite{beane,ioffe}. 
It is the purpose of this note, to
examine this point by studying the consequences of the chiral symmetry limit
for the decay amplitude. We will show, that chiral symmetry,
broken or unbroken, does not forbid the decay for the model considered before, in which the
two scalar diquarks are in a relative $P$-wave state. On the other hand, for a pentaquark with
an internal structure which may be viewed as an antiquark and two chirally different 
diquarks in a relative $S$-wave, strictly unbroken chiral symmetry leads to a
vanishing decay amplitude. 
However, because chiral symmetry is spontaneously broken by quark
condensates this stability is lost. Moreover, the large magnitude and 
spatial extension of the nonlocal quark condensate leads also in this
case to a large decay width if not reduced by a small wave function 
overlap in coordinate space.

\section{Chiral symmetry for nucleons and pentaquarks}
The spontaneously broken chiral symmetry does not leave the vacuum
invariant. The axial charges generate the Goldstone particles. 
Thus, chiral symmetry is not a symmetry of the particle spectrum. However, 
it can be used for selecting the couplings of fields in effective
Lagrangians. We therefore consider interpolating fields for the usual 
baryon octet and for the pentaquark and study their possible couplings.

We start by defining Lorenz scalar left and right diquark
field operators and state their $SU(3)_L\times SU(3)_R$ 
representations of the chiral group.
Using the convention that the quark fields $q_L$ and $q_R$ transform
according to (3,1) and (1,3), respectively, one finds 
\begin{eqnarray}
D_L^{\alpha,i} &=& 
\epsilon^{\alpha\beta\nu} 
\epsilon^{ijk}
\left((q_L)^T_{\beta,j} 
\gamma_5 C (q_L)_{\nu,k}\right), 
\nonumber\\
D_R^{\alpha,i} &=& 
\epsilon^{\alpha\beta\nu} 
\epsilon^{ijk}
\left((q_R)^T_{\beta,j} 
\gamma_5 C (q_R)_{\nu,k}\right).
\end{eqnarray}
Here, $\alpha,\beta,\gamma$ are color indices and $C=-i\gamma_0\gamma_2$ is the charge conjugation matrix.
We note that left- and right-handed scalar diquark fields are in
irreducible representation of the chiral group:  
$D_L$ transforms according to $(\bar 3,1)$ and 
$D_R$ according to $(1,\bar 3)$.

Without loss of generality the state vectors of the baryon octet
which contains proton and neutron formed by 3 quarks can be written in a Fock space representation as a quark-diquark
combination \cite{dosch}. It is then suggestive to use among the possibilities
for baryon field operators an equivalent quark-diquark
form which can generate these baryons 
\begin{eqnarray}
\label{bb}
B_i^j  =  \mbox{$\frac{1}{2}$}\{D_L^{\alpha,j} +D_R^{\alpha,j}  - \gamma_5
(D_L^{\alpha,j} -D_R^{\alpha,j} )\} \;q_{\alpha,i}.
\end{eqnarray}
The baryon field is written in such a way that the left and right handed
components are in left and right  representations of the chiral group: 
\begin{eqnarray}
\label{form1}
(B_L)_i^j  =   D_L^{\alpha,j}(q_L)_{\alpha,i}, \quad  
(B_R)_i^j  = D_R^{\alpha,j}(q_R)_{\alpha,i}.
\end{eqnarray}
$B_L$ transforms as (1+8,~1), and $B_R$ as (1~,1+8). 
We can also get left and right baryon field components in
the different form
\begin{eqnarray}
\label{form2}
(B_L)_i^j  =   D_R^{\alpha,j}(q_L)_{\alpha,i}, \quad  
(B_R)_i^j = D_L^{\alpha,j}(q_R)_{\alpha,i}.
\end{eqnarray}
by simply reversing the (negative) sign for the $\gamma_5$ term in (\ref{bb}).
Then $B_L$ transforms as $(3,\bar 3)$ and $B_R$ as $(\bar 3,3)$.
Since the gluon coupling in QCD is helicity conserving, the attraction
between quark and diquark favours the form (\ref{form1}). 
Nevertheless, the latter case (\ref{form2}) will also be considered below.

\subsection{Pentaquarks with an internal $P$-wave structure} 
For the interpolating pentaquark field  with two scalar diquarks in a relative $P$-wave state we
take again an expression which leads to separate left and right
representations for $P_L$ and $P_R$:
\begin{eqnarray}
\label {P}
P^{abc} &= &  \mbox{$\frac{1}{2}$}\epsilon_{\alpha\beta\nu}
[( D_L^{\alpha, a}
\stackrel{\leftrightarrow}{ \partial^\mu}
D_L^{\beta,b}+ D_R^{\alpha, a} \stackrel{\leftrightarrow}{ \partial^\mu}D_R^{\beta, b}) 
\nonumber
\\
&-&
\gamma_5(D_L^{\alpha, a}\stackrel{\leftrightarrow}{\partial^\mu}
D_L^{\beta, b}- D_R^{\alpha, a} \stackrel{\leftrightarrow}{ \partial^\mu}
D_R^{\beta,b})]\gamma_5\gamma_\mu C (\bar q^T)^{\nu,c} 
\nonumber\\
P_L^{abc} &=& \epsilon_{\alpha\beta\nu}D_L^{\alpha, a}  
\stackrel{\leftrightarrow}{ \partial^\mu}  D_L^{\beta, b} 
\sigma_\mu i\sigma_2 (q^{*}_L)^{\nu, c}
 \nonumber\\
P_R^{abc} &=& \epsilon_{\alpha\beta\nu}~ D_R^{\alpha, a}  
\stackrel{\leftrightarrow}{ \partial^\mu}  D_R^{\beta, b} 
\bar \sigma_\mu i\sigma_2 (q^*_R)^{\nu, c}.
\end{eqnarray}
Symmetrization with respect to the indices $a$ and $b$ is implied. In (\ref{P}) $P_L, P_R$ and $q_L,q_R$ denote  
two-component Weyl fields transforming as left and right handed  spinor fields, respectively. 
Obviously, $P$ contains the antidecuplet (with respect to the diagonal subgroup $SU(3)_V$) we are interested in and can be extracted from it. The parity of $P$ is even. The expression for $P$
includes (covariant) derivatives to represent the  relative 
$P$-wave state of the two diquarks in a local form \cite{eidemuller}. The local form is convenient but not necessary. Chiral transformations are global transformations. It is evident that
$P_L$ transforms as $(8+\overline{10}, 1 )$ and $P_R$ as $(1,8+\overline{10})$. Consistent with the chosen form
for the baryon octet all quarks in $P_L$ are lefthanded and all
quarks forming $P_R$ are right handed.

The combination $ B^{\dagger}_L \overline\sigma_\mu  P_L$, 
can form a left-handed octet current transforming as $(8,1)$ when applying proper 
Clebsch-Gordan coefficients. 
This is evident from the transformation properties given above.
Similarly,
$B^{\dagger}_R  \sigma_\mu P_R$
can form a right-handed current transforming as $(1,8)$. Together, these combinations have the correct properties of  an  axial vector octet and therefore can couple to an axial vector field with a 
chirally invariant coupling constant!

We can also construct a chirally invariant coupling of this "$P$-wave
pentaquark" to the baryon octet and the Goldstone particles. For this
purpose we use
the non-linear representation for the light meson octet
\begin{eqnarray}
\label{sigma}
\Sigma  =  \exp \left(i \lambda_k \phi^k/{F_\pi}\right)
\end{eqnarray}
with the unitary matrix $\Sigma$ transforming according to $(3,\bar 3)$.
Here $\lambda^k$ denote the Gell-Mann matrices and $\phi_k$ the 8 pseudoscalar 
meson fields. Obviously, chiral transformations of $\Sigma$ leave
this matrix unitary and  define thereby the transformation properties of the
Goldstone fields. For an earlier application of this matrix (on the proton spin problem) see \cite{ellw}. 
Let us now introduce the Dirac matrices
\begin{eqnarray}
\ K_L^\zeta &=&  \gamma^\mu \mbox{$\frac{1}{2}$}(1-\gamma_5)
{\rm Sp}(\Sigma\stackrel{\leftrightarrow}{\partial^\mu} \Sigma^\dagger~ 
\lambda^\zeta),
\nonumber
\\
K_R^\zeta &=&   \gamma^\mu \mbox{$\frac{1}{2}$}  
(1+\gamma_5){\rm Sp}(\Sigma^\dagger  \stackrel{\leftrightarrow}{ \partial^\mu}
\Sigma\lambda^\zeta), 
\end{eqnarray}
and the Clebsch-Gordan coefficients
\begin{eqnarray}
G_\rho^{\zeta,\xi}= 
\langle\overline{10}_\rho|8_\zeta,8_\xi\rangle, \quad 
\rho= 1\dots 10, \quad \zeta, \xi=1\dots 8.
\end{eqnarray}
The  baryon field can now
be coupled to the  pentaquark field in a chiral invariant way. For the left 
handed part one has
\begin{eqnarray}
\left(\overline{B^\xi}~ G^\rho_{\xi,\zeta} K_L^\zeta ~(P)_\rho\right).
\end{eqnarray}
The corresponding right-handed coupling reads
\begin{eqnarray}
\left(\overline{B^\xi}~ G^\rho_{\xi,\zeta} K_R^\zeta~ (P)_\rho\right).
\end{eqnarray}
Adding both expressions and expanding $\Sigma$ to first order in
the Goldstone
fields gives now the coupling of the pentaquark decuplet to the
baryon octet and the Goldstone particles. Chiral symmetry allows this 
derivative coupling. 
There exists no symmetry
argument for the corresponding coupling constant to vanish. This
result is in agreement with the numerical values for the width
obtained in \cite{mss} which turned out to be generally large. Only a
small spatial overlap between this pentaquark and the nucleon can
reduce the width. The existence of a chirally invariant coupling for
this pentaquark can be traced back to the fact that $\gamma_\mu C \bar
q^T$ transforms like the diquarks. Without the $\gamma_\mu $ term and thus without 
the $P$-wave structure the result will be quite different.

\subsection{Pentaquarks stable in the strict chiral symmetry limit}
Therefore, let us now consider an interpolating pentaquark field
of positive parity which generates the two diquarks in an $S$-wave state:
\begin{eqnarray}
\label{ps}
P^{abc} &=&  \mbox{$\frac{1}{2}$}\epsilon_{\alpha\beta\gamma}[(D_L^{\alpha, a}
D_R^{\beta,b}+D_R^{\alpha, a}  D_L^{\beta, b}
\nonumber\\
&&- \gamma_5 (D_L^{\alpha, a}
D_R^{\beta,b}-D_R^{\alpha, a}  D_L^{\beta,b})]~\gamma_5 C~  (\bar q^T)^{\gamma,c}.
\nonumber\\
\end{eqnarray}
In the two-component Weyl field representation  we have
\begin{eqnarray}
\label{PLR}
(P_L)^{abc}  &=&  \epsilon_{\alpha\beta\gamma}~D_L^{\alpha, a}  D_R^{\beta, b}  
i \sigma_2 (q_R^*)^{\nu,c},
\nonumber
\\
(P_R)^{abc}  &=&   \epsilon_{\alpha\beta\gamma}~D_R^{\alpha, a}  D_L^{\beta, b} 
i \sigma_2 (q_L^*)^{\nu,c}.  
\end{eqnarray} 
$P_L$ transforms as $(\bar 3 , 3 + \bar 6)$ and $P_R$ as $(3 + \bar 6 ,  \bar 3)$. 
The pentaquark $SU(3)_V$ antidecuplet arises  from the $\bar 6$
content in these expressions.
The interpolating field operator which generates "$S$-wave pentaquarks" 
of negative parity is obtained by multiplying (\ref{ps})
by $\gamma_5$. Evidently, this negative parity pentaquark has the same 
transformation properties under chiral transformation as the one with 
positive parity.

As it is seen from these transformation properties, this time 
the left-handed axial current formed from 
$B_L^\dagger \overline \sigma_\mu P_L$ transforms as 
$(\bar 3+6 +15,3+\bar 6)$ and not as $(8,1)$ as required. 
This is in strong contrast to the case of the
"$P$-wave" type pentaquark we discussed before. A similar result 
holds if we take for the baryon the form in which $B_L$ transforms 
as $(3,\bar 3)$ and $B_R$ as $(\bar 3,3)$. It implies, that in the 
strict chiral symmetry limit the desired axial vector current cannot be 
constructed and this pentaquark is stable. The vanishing of the decay 
amplitude of "$S$-wave pentaquarks" in case of an unbroken chiral symmetry
is in accord with the findings of Ioffe and Oganesian \cite{ioffe}. 
The $P$-wave pentaquark and the positive parity $S$-wave pentaquark 
have identical quantum numbers: 
total angular momentum, $SU(3)$ quantum numbers and parity. 
But they differ in their chiral transformation properties and their 
$\gamma_5$ symmetry\footnote{This 
$\gamma_5$ operation reminds us of the very first proposals for a 
$\gamma_5$ symmetry for all 
particle interactions: The invariance of the weak interaction with 
regard to a $\gamma_5$ operation on the elementary particles at the time 
was postulated in \cite{stech_jensen}. That also the strong interactions  
are invariant under 
this symmetry was suggested in \cite{rollnik} and accomplished for a 
non linear $\pi$ meson 
nucleon interaction.} (a discrete subgroup of $U(3)_L \times U(3)_R$): 
multiplying all quark fields by $\gamma_5$ gives 
$+ \gamma_5$ for the "$P$-wave pentaquark" and the nucleon octet 
but $-\gamma_5$ 
for the $S$-wave pentaquark. The axial vector constructed from the 
baryon octet and this 
latter pentaquark then changes sign under this transformation \cite{ioffe}, 
another reason 
for a vanishing coupling to the axial field in the strict chiral limit. 
In this limit a coupling to the pseudoscalar meson nonet is not possible. 
According to the derivation it is clear that the precise internal structure 
of the pentaquark is not essential, 
only the transformation properties matter: The diquarks do not have to be 
of a small 
size and may strongly overlap with each other and the antiquark. 

\subsection{The spontaneous breaking of chiral symmetry}

The spontaneous breaking of chiral symmetry changes the situation. 
Quark condensates appear and the quantity $\Sigma$, the
non-linear form for the light meson octet  given in (\ref{sigma}), can now be 
used to change the transformation properties of the fields: 
$\Sigma ~ q_R$ transforms as $q_L$, and $\Sigma^\dagger ~ q_L$ 
transforms as $q_R$. An appropriate application of $\Sigma$
allows the non vanishing of the axial vector matrix element for the transition 
to nucleons.
Thus, because of the spontaneous symmetry breaking also the $S$-wave  
pentaquark looses its stability.
In a Fock space model where all quark fields act on the vacuum at equal 
times (or on a light like hyper plane) but at different positions it is 
easy to see the reason for the stability of the $S$-wave pentaquark in 
case of the strict chiral symmetry, and for its instability due to the 
spontaneous chiral symmetry breaking: In the unbroken case the axial vector 
current matrix element for the transition amplitude can be calculated by 
commuting the fields using the equal time commutation relations. As it is 
obvious from (\ref{form1}) and (\ref{PLR}) this gives zero for our 
$ S$-wave pentaquark since $q_L$ commutes with $q^{\dagger}_R$ and $q_R$ with
$q^{\dagger}_L$ . However, since chiral symmetry is spontaneously broken, 
non local condensates such as $\langle \psi_L(x) \psi^{\dagger}_R(0)\rangle|_{x_0=0} $ 
survive. (The correct gauge invariant form for these condensates includes 
a Schwinger string not  shown here.) These condensates replace the 
$\delta^3(x)$ function obtained from equal time commutators in transitions 
which are not suppressed by chiral symmetry. One can compare now  the space 
integral of the condensate with the space integral over the  $\delta^3(x)$ 
function (which is 1). This gives a measure of the importance of the 
spontaneous symmetry breaking. We take for the condensate the fit formula 
given in the literature \cite{rad,doja}.
\begin{equation}
\langle\psi_L(x_0=0,\vec x) \psi^{\dagger}_R (0)\rangle= 
\frac{1}{2} \langle \bar \psi \psi \rangle e^{-\vec x^2 M^2_0/16}
\end{equation}
With $\langle\bar \psi \psi\rangle \approx (254~{\rm MeV})^3$ and $M_0
\approx 860~{\rm MeV}$	 
the numerical value of the space integral turns out to be $\approx 4.6$.
In an actual calculation of the transition amplitude this space
integral will be somewhat reduced by the variation of the wave
function multiplying the condensate, but it will certainly stay of
order one. Consequently, we are forced to conclude: In spite of
the vanishing of the decay amplitude in the unbroken chiral symmetry
limit, the spontaneous breaking of this symmetry is strong and leads 
in general to amplitudes comparable with the ones which are not
inhibited by the unbroken symmetry.
It appears, that chiral symmetry cannot explain the small width of 
pentaquarks except for the improbable case that the spontaneous  
breaking of chiral symmetry is less effective in the five quark
system. 
If pentaquarks exist, their small widths are likely caused by an 
unusual spatial structure of these particles leading to a small wave 
function overlap with the nucleon wave function.

\section{Summary}
In conclusion, we can say that for the case of a pentaquark with a
$P$-wave internal structure a direct coupling between the pentaquark 
antidecuplet, the Goldstone particles, and the baryon octet 
is allowed in the limit of an unbroken chiral symmetry. Thus, the decay 
amplitude for a $P$-wave pentaquark does not vanish in the chiral
limit. In general, this leads to a large decay width which can only be 
numerically suppressed by a small spatial overlap of this pentaquark 
state with the nucleon. 

On the other hand, a pentaquark with an internal $S$-wave structure 
(essentially an $S$-wave between two diquarks of different chirality)
becomes stable in the limit of strict chiral symmetry. However, it can
decay by emitting a Goldstone particle because of the spontaneous
breaking of chiral symmetry.  
It turns out that this breaking, which is caused by quark condensates,
is strong and washes out the inhibition of the unbroken symmetry. 
Our results imply that the small width of the pentaquark is not caused
by chiral symmetry effects, it must have a different origin.

\acknowledgments  We thank Bob Jaffe, Holger Gies, Otto Nachtmann, 
Matthias Jamin, Silvano Simula and in particular Heiri Leutwyler for
interesting discussions and communications. 
This work was supported by the Alexander von Humboldt-Stiftung in form 
of a "R\"uckkehr Stipendium" for D.M.

\end{document}